\input ./style/arxiv-general.cfg
\documentclass[MSNbibl,number,citesort,dvips]{arxbj}
\makeatletter
   \@ifpackageloaded{graphicx}{}{\usepackage{graphicx}}
\makeatother
\usepackage{algorithm,algorithmic}
\usepackage{upgreek}
\usepackage{mathrsfs}

\volume{22}
\issue{2}
\pubyear{2016}
\firstpage{901}
\lastpage{926}
\doi{10.3150/14-BEJ681}
\docsubty{FLA}

\makeatletter
\def\mid{|}
\def\e{\mathrm{e}}
\newcommand{\eqref}[1]{(\ref{#1})}
\def\B{\mathcal{B}}
\def\D{\mathscr{D}}
\def\E{\mathbb{E}}
\def\P{\mathscr{P}}
\def\R{\mathbb{R}}
\def\X{\mathbb{X}}

\newcommand{\cl}[1]{\mathcal{#1}}
\newcommand{\bb}[1]{\mathbb{#1}}
\newcommand{\msf}[1]{{#1}}

\newtheorem{theorem1}{Theorem}[section]
\newproclaim{assumption}{Assumption}
\newtheorem{proposition}[theorem1]{Proposition}
\newproclaim{definition}[theorem1]{Definition}

\newremark{remark1}[theorem1]{Remark}
\newremark{example1}[theorem1]{Example}

\makeatother

\begin{document}
\begin{frontmatter}

\title{Dynamic density estimation with diffusive Dirichlet mixtures}

\runtitle{Dynamic estimation with diffusive mixtures}

\begin{aug}
\author[A]{\inits{R.H.}\fnms{Rams\'es H.}~\snm{Mena}\thanksref{A}\ead[label=e1]{ramses@sigma.iimas.unam.mx}}
\and
\author[B]{\inits{M.}\fnms{Matteo}~\snm{Ruggiero}\corref{}\thanksref{B}\ead[label=e2]{matteo.ruggiero@unito.it}}
\address[A]{Departamento de Probabilidad y Estad\'istica, IIMAS, UNAM,
Apartado Postal 20-126, 01000, M\'exico, D.F. Mexico.
\printead{e1}}

\address[B]{Collegio Carlo Alberto \& ESOMAS Department,
University of Torino, C.so Unione Sovietica 220, 10134, Torino, Italy.
\printead{e2}}
\end{aug}

%
\received{\smonth{9} \syear{2013}}
%
\revised{\smonth{5} \syear{2014}}

%
\begin{abstract}
We introduce a new class of {nonparametric} prior distributions on the
space of continuously varying densities, induced by Dirichlet process
mixtures which diffuse in time. These select time-indexed random
functions without jumps, whose sections are continuous or discrete
distributions depending on the choice of kernel. The construction
exploits the widely used stick-breaking representation of the Dirichlet
process and induces the time dependence by replacing the stick-breaking
components with one-dimensional Wright--Fisher diffusions. These
features combine appealing properties of the model, inherited from the
Wright--Fisher diffusions and the Dirichlet mixture structure, with
great flexibility and tractability for posterior computation. The
construction can be easily extended to multi-parameter GEM marginal
states, which include, for example, the Pitman--Yor process. A full
inferential strategy is detailed and illustrated on simulated and real data.
\end{abstract}

%
\end{frontmatter}

\section{Introduction}\label{sec1}

Bayesian nonparametric inference has undergone a tremendous development
in the last two decades. This has been stimulated not only by
significant theoretical advances, but also by the availability of new
and efficient computational methods that have made inference, based on
analytically intractable posterior distributions, feasible. A recent
extensive survey of the state-of-the-art of the discipline can be found
in Hjort, Holmes, M\"uller and Walker \cite{HHMW}.

In this paper, we tackle the problem of estimating continuously varying
distributions, given data points observed at different time intervals,
possibly not equally spaced. More specifically, we consider the
following setting, stated as an assumption for ease of later reference.

\begin{assumption}\label{assumption}
The data generating process is assumed to be a random function $g\dvtx \X
\times[0,T]\rightarrow\R_{+}$, $0<T< \infty$, where $\X$ is a locally
compact Polish space, $g$ is continuous in the sense that $\sup_{|s-t|<\delta}\sup_{x\in\X}|g(x,s)-g(x,t)|\to0$ as $\delta\to
0$, and
sections $g(\cdot,t)$ are densities absolutely continuous with respect
to a common dominating measure.
\end{assumption}

Here we assume that given $g(\cdot,t)$, the data are such that at time $t_{i}$
%
\begin{equation}
\label{data} y_{t_{i,1}},\ldots,y_{t_{i,k}}\stackrel{\mathrm{i.i.d.}} {\sim}g(
\cdot,t_{i}),
\end{equation}
independently of the past observations $y_{t_{j},h}$, for all $j<i$ and
all $h$. Hence, observations $y_{t_{i},j}$ are exchangeable for fixed
$i$ and varying $j$, but only partially exchangeable in general.
Our goal is to define an appropriate prior for $g$ and carry out
inference on its entire shape.
We will deal with both single and multiple data points available at
every time $t_{i}$, obtained by letting $k=1$ and $k>1$ in \eqref
{data}, and refer to these two settings as \emph{single data} and
\emph
{multiple data}, respectively. These are of separate interest as in
some frameworks only single data points are available (e.g., financial
indices); it is then natural to wonder about the model performance in
such settings, with structural lack of information, while evaluating
the precision gain that can be obtained when more information is available.

The commonly recognised cornerstone for density estimation in a
Bayesian nonparametric framework is the Dirichlet process mixture
model, introduced by Lo \cite{L84} and recalled in Section~\ref{secdepprocesses} below.
Here the idea is to extend {some} advantages of using
a Dirichlet process mixture to the case when $g$ is considered to be
the realisation of a measure-valued process. Hence, we aim at inducing
a prior on the space of functions as in Assumption~\ref{assumption} by
constructing a measure-valued process, with continuous sample paths and
marginal states given by a Dirichlet process mixture, which is suitable
for nonparametric inference in continuous time. Different types of
temporal dependence induced on the observations are of interest. Here
we focus on Markovianity for the mixing measure. Although potential
applications of assuming a Markovian mixing measure concern a variety
of fields such as econometrics, finance and medicine among others, this
approach has been object of only a limited amount of research, so far,
in the literature on Bayesian nonparametric inference. On the other
hand, in a context of temporal dependence, the Markov property for the
mixing measure considerably simplifies the finite dimensional
distributions structure, and, most importantly, it need not impose a
Markovian dependence on the actual observations.

Thus, our setting can be regarded in two ways. One is that of a
nonparametric regression, where the mixing measure is indexed by the
covariate $t$, and the observations are partially exchangeable. A
second interpretation, given our above requests, is that of a hidden
Markov model, whereby the unobserved signal is an infinite-dimensional,
or measure-valued, Markov process. Conditionally on the knowledge of
the signal $g(\cdot,t)$, the observations are independent of each
other, and on the past observations, with emission distribution given
by the signal state, as in \eqref{data}.

The paper is organized as follows. Section~\ref{secdepprocesses}
recalls the essential preliminary notions, like that of Dirichlet
process and Dirichlet process mixture, and reviews to some extent the
literature on the so-called dependent processes in Bayesian
nonparametrics. These are in fact measure-valued processes, indexed by
time or more generally by covariates, specifically designed for
inferential purposes in Bayesian nonparametric frameworks with
nonexchangeable data.
Section~\ref{secDPM} moves from the stick-breaking representation of
the Dirichlet process to construct a class of diffusive Dirichlet
processes. The idea consists in replacing the Beta-distributed
components of the stick-breaking weights with one-dimensional
Wright--Fisher diffusions. This yields a time-dependent process with
purely-atomic continuous paths and marginal states given by Dirichlet
processes. In order to have a statistical model suitable for estimating
functions as in Assumption~\ref{assumption}, we then define a diffusive
Dirichlet mixture by considering an appropriate hierarchy whose top
level is given by a diffusive Dirichlet process.

A challenging aspect of statistical models which involve diffusions
regards the computational side, even when this is performed via
simulation techniques with the aid of a computer. In a nutshell, this
is due to the {often encountered} intractability of the transition
density, in the fortunate cases when this is known explicitly.
Here we devise a strategy for posterior computation based on Gibbs
sampling and slice sampling, with the latter used both on the
instant-wise infinite-dimensional mixing measure and on the transition
density of the time-dependent components. After outlining the algorithm
for posterior computation in Section~\ref{postcomputation}, in Section~\ref{secillustration} we illustrate the use of diffusive Dirichlet
mixtures on {two sets of} simulated data and on real financial data.
Section~\ref{secdiscussion} collects some concluding remarks and
briefly highlights possible extensions, concerned with the model
parametrisation and with the aim of relaxing some model constraints.
All proofs and the algorithm details are deferred to the \hyperref[app]{Appendix}.


\section{Dependent processes in Bayesian nonparametrics}\label{secdepprocesses}

The Dirichlet process, introduced by Ferguson \cite{F73} and widely
accepted as
the main breakthrough {in the history of} Bayesian nonparametric
statistics, is a discrete random probability measure defined as
follows. Let $\X$ be a Polish space endowed with the Borel sigma
algebra $\B(\X)$, $\P(\X)$ be the space of Borel probability measures
on $\X$, and let $\alpha$ be a nonatomic, finite and nonnull measure
on $\X$. A $\P(\X)$-valued random variable $Q$ is said to be a
Dirichlet process with parameter $\alpha$, denoted $Q\sim\D_{\alpha}$,
if for any measurable partition $A_{1},\ldots,A_{k}$ of $\X$, the
vector $(Q(A_{1}),\ldots,Q(A_{k}))$ has the Dirichlet distribution with
parameters $(\alpha(A_{1}),\ldots,\alpha(A_{k}))$.
A~second construction of the Dirichlet process, still due to Ferguson
\cite
{F73}, exploits the idea of normalising the jumps of a gamma
subordinator by their sum, locating the jumps at independent and
identically distributed points sampled from $\alpha/\alpha(\X)$. This
strategy has been followed for constructing several other nonparametric
priors, among which the normalised inverse-Gaussian process Lijoi, Mena
and Pr\"unster \cite
{LMP05} and the normalised generalized gamma process Lijoi, Mena and
Pr\"unster \cite{LMP07}.
A later construction of the Dirichlet process, formalized by Sethuraman~\cite{S94}
and particularly useful for our purposes, is usually referred to as the
\emph{stick-breaking representation}. This states that the law of a
Dirichlet process coincides with the law of the discrete random
probability measure
%
\begin{equation}
\label{RPM-series} S=\sum_{i=1}^{\infty}w_{i}
\delta_{x_{i}},
\end{equation}
obtained by letting
%
\begin{equation}
\label{stick-breaking-weights} w_{1}=v_{1},\qquad w_{i}=v_{i}
\prod_{j=1}^{i-1}(1-v_{j}),\qquad
v_{i}\stackrel{\mathrm{i.i.d.}} {\sim} \operatorname{Beta}(1,\theta),
\end{equation}
and $x_{i}\sim^{\mathrm{i.i.d.}}\alpha/\theta$, where $\theta=\alpha(\X)$
{and the
$v_{i}$'s and $x_{i}$'s are mutually independent}. The stick-breaking
construction has received a wide appreciation from the Bayesian
community. In particular, this is due to the fact that {it greatly
facilitates the implementation of posterior simulation, using} Markov
chain Monte Carlo techniques that exploit the slice sampler (Damien,
Wakefield and Walker \cite
{DWW99}, Walker \cite{W07}) {or} the retrospective sampler
(Papaspiliopoulos and Roberts \cite{PR08}). Such
strategies for avoiding infinite computations, without
deterministically truncating the random measure, can be used, for
example, with Dirichlet process mixtures (Lo \cite{L84}). The latter are a
very popular class of models which amplifies the use of Dirichlet
processes to a wider spectrum of statistical applications, most notably
density estimation and data clustering, by modelling observations
according to the random density
%
\begin{equation}
\label{mixturemodel} f_{S}(y)=\int K(y| x)S(\mathrm{ d}x)=\sum
_{i=1}^{\infty}w_{i}K(y| x_{i}),
\end{equation}
where $K(\cdot|y)$ is a kernel density and $S$ is as in (\ref
{RPM-series}). {Thanks to the large support properties of the Dirichlet
prior, the above model results in a very flexible class of
distributions: for instance, any density on the real line can be
recovered as an appropriate Dirichlet mixture of normal densities. For
this and other examples see Lo \cite{L84}, Section~3 and Ghosh and
Ramamoorthi \cite{GR03},
Section~5.}

Many developments which {derive from this modelling approach} have been
proposed. A first possibility is to replace the Dirichlet process in
(\ref{mixturemodel}) by letting $S$ be some other discrete nonparametric prior. See Lijoi and Pr\"unster \cite{LP10} and references therein.
A different direction, which currently represents a major {research
frontier} of the area, is to extend $S$ in (\ref{mixturemodel}) in
order to accommodate forms of dependence more general than exchangeability.
{Besides pioneering contributions stimulated by Cifarelli and Regazzini
\cite{CR78},} this line
of research was initiated by MacEachern \cite{ME99,ME00}, who proposed
a class of \emph{dependent processes}, that is a collection of random
probability measures
%
\begin{equation}
\label{depprocesses} \Biggl\{S_{u}=\sum_{i=1}^{\infty}w_{i}(u)
\delta_{x_{i}(u)}, u\in \mathbb {U} \Biggr\},
\end{equation}
where the weights $w_{i}$ and/or the atoms $x_{i}$ depend on some
covariate $u\in\mathbb{U}$.
The current literature on the topic includes, {among other contributions}:
De Iorio, M\"uller, Rosner and MacEachern~\cite{DIetal04}, who
proposed a model with an ANOVA-type dependence structure;
Gelfand, Kottas and MacEachern \cite{GKM05}, who apply the dependent
Dirichlet process to spatial
modelling by using a Gaussian process for the atoms;
Griffin and Steel \cite{GS06}, who let the dependence on the random
masses be directed
by a Poisson process;
Dunson and Park \cite{DP08}, who construct an uncountable collection
of dependent
measures based on a stick-breaking procedure with kernel-based weights;
Rodriguez and Dunson \cite{RD11}, who replace the Beta random variables
in (\ref
{stick-breaking-weights}) with a transformation of Gaussian processes
via probit links.
See also
Dunson, Pillai and Park \cite{DPP07}, who define a Dirichlet mixture
of regression models;
Dunson, Xue and Carin \cite{DXC08}, who propose a matrix-valued
stick-breaking prior;
Duan, Guindani and Gelfand \cite{DGG07} and Petrone, Guindani and
Gelfand \cite{PGG09} for other developments of dependent
priors for functional data;
Fuentes-Garc\'ia, Mena and Walker \cite{FMW09} for a dependent prior
for density regression;
{Trippa, M\"uller and Johnson~\cite{TMJ11}, who define a dependent
process with Beta marginals.}

Of particular interest for the purposes of this paper are the
developments of dependent processes where the space $\mathbb{U}$
indexes time. In this regard we mention, among others,
Dunson \cite{D06}, who models the dependent process as an
autoregression with
Dirichlet distributed innovations, whereas in Griffin and Steel \cite
{GS10} the
innovation is reduced to a single atom sampled from the centering measure;
Caron \textit{et al.} \cite{Cea06}, who model the noise in a dynamic linear
model with a
Dirichlet process mixture;
Caron, Davy and Doucet \cite{CDD07}, who develop a time-varying
Dirichlet mixture with
reweighing and movement of atoms;
Rodriguez and Ter Horst \cite{RT08}, who induce the dependence in time
only via the atoms, by
making them into an heteroskedastic random walk.

Here we aim at constructing a measure-valued diffusion whose
realisations are functions as in Assumption~\ref{assumption}. In this
respect, the infinite-dimensional diffusions, found in the literature,
which are related to Bayesian nonparametric priors are not suitable for
efficient inference. Just to mention a few cases, this holds, for
example, for the infinitely-many-alleles model (Ethier and Kurtz \cite
{EK81,EK86}),
related to the Dirichlet process; for its two-parameter version (Petrov
\cite
{P09}), related to the two-parameter Poisson--Dirichlet distribution;
and for normalised inverse-Gaussian diffusions (Ruggiero, Walker and
Favaro \cite{RWF13}), related
to normalised inverse-Gaussian random measures. The reasons are mainly
due to the fact that an inferential strategy based on these dependent
processes, given our current knowledge of their properties, would
oblige to update single atoms of a P\'olya urn scheme (see, e.g.,
Ruggiero and Walker \cite{RW09}; Favaro, Ruggiero and Walker \cite
{FRW09}), or otherwise face serious
computational issues.
To the best of our knowledge, the only model which satisfies the given
requirements, {among which the Feller property,} and allows efficient
inference is given in Mena, Ruggiero and Walker \cite{MRW11}. However,
this is based on
decreasingly ordered weights, so, despite showing good performance,
this feature can nonetheless be considered restrictive for certain
applications. The model developed in the next section removes this
constraint by letting only the weights' means be ordered, as happens
for the Dirichlet process.


\section{Diffusive Dirichlet process mixtures}\label{secDPM}

\begin{figure}

\includegraphics{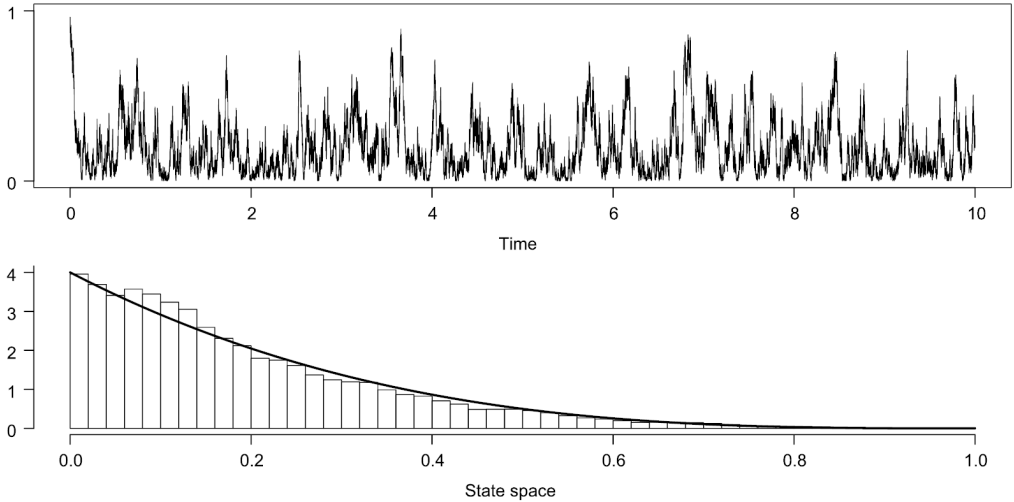}

\caption{Above: an approximated sample-path of a
Wright--Fisher diffusion, with $a=1$ and $b=4$. Below: ergodic
frequencies of the above sample-path against invariant distribution
$\operatorname{Beta}(1,4)$.}\label{figWF}
\end{figure}

{In this section, we elaborate on a construction provided in Feng and
Wang \cite
{FW07}, in order to develop a class of measure-valued processes,
suitable to be used in a statistical model, with the sought-after
features outlined in the \hyperref[sec1]{Introduction}.}
Consider the special case of (\ref{depprocesses}) given by the
collection of random probability measures
%
\begin{equation}
\label{dependentprocess} P_{t}=\sum_{i=1}^{\infty}w_{i}(t)
\delta_{x_{i}},\qquad t\ge0, x_{i}\stackrel{\mathrm{i.i.d.}} {\sim}G,
\end{equation}
where $\sum_{i\ge1}w_{i}(t)=1$ for all $t\ge0$ and $G$ is a nonatomic
probability measure on $\X$. Here the atoms are random but do not vary
with time, and the dependence is induced only via the weights. This
setting suffices for guaranteeing enough modelling flexibility (see
Section~\ref{secdiscussion} for more comments on this point), whereas
the opposite scheme, obtained by inducing the dependence only via the
atoms, may be unsatisfactory. See, for example, Rodriguez and Dunson
\cite{RD11} for a discussion.
Recall now the stick-breaking structure of the Dirichlet process
weights \eqref{stick-breaking-weights}. A natural extension for having
time dependence with continuity in $t$ is to let $w_{i}$ diffuse in
time, in a way that retains the marginal distributions. A simple way to
achieve such result is to let each component $v_{i}$ diffuse in
$[0,1]$, with fixed marginals, and perform the same construction as in
\eqref{stick-breaking-weights}. This can be obtained by letting the
$v_{i}$'s be one-dimensional Wright--Fisher diffusions, characterised
as the unique solution in $[0,1]$ of the stochastic differential equation
%
\begin{equation}
\label{wf-sde} \mathrm{ d}v(t)=\tfrac{1}{2}\bigl[a\bigl(1-v(t)\bigr)-bv(t)
\bigr]\,\mathrm{ d}t +\sqrt{v(t) \bigl(1-v(t)\bigr)} \,\mathrm{ d}B(t),
\end{equation}
where $a,b\ge0$ and $B(t)$ denotes a standard Brownian motion.
See, for example, Karlin and Taylor \cite{KT81}, Section~15.2.
For our purposes, it will suffice to highlight the following properties
of Wright--Fisher diffusions $v(t)$ with parameters $(a,b)$ as in
\eqref
{wf-sde}, henceforth denoted $v(\cdot)\sim\operatorname{WF}(a,b)$:
\begin{itemize}[$-$]
\item[$-$] when $v(t)$ approaches 0 (resp., 1), the diffusion coefficient
converges to 0 and the drift approaches $a/2$ (resp., $-b/2$), thus
keeping the diffusion inside $[0,1]$;
%
\item[$-$] when $a,b\ge1$, the points 0 and 1 are both \emph{entrance
boundaries}, implying (essentially) that they are never touched for $t>0$;
\item[$-$] when $a,b>0$, $v(t)$ has invariant distribution given by a
$\operatorname
{Beta}(a,b)$;
\item[$-$] when $a,b>0$, $v(t)$ is strongly ergodic, that is, irrespective
of the initial distribution, the law of $v(t)$ will converge to the
invariant measure {as $t$ diverges}.
\end{itemize}
A typical behavior of $v(t)$, together with the occupancy frequencies
plotted against the invariant distribution, is shown in Figure~\ref{figWF}, for values $(a,b)=(1,4)$. Note how these parameters make the
trajectory occupy the half interval containing the mean value of $\operatorname{Beta}(1,4)$ for most of the time.
%

The idea is then to let every $v_{i}$ in (\ref{stick-breaking-weights})
vary, {independently of the other components}, according to a
Wright--Fisher diffusion with parameters $(1,\theta)$, that is
$v_{i}(\cdot)\sim^{\mathrm{i.i.d.}} \operatorname{WF}(1,\theta)$, with
$v_{i}(\cdot)=\{
v_{i}(t),t\ge0\}$, and to construct diffusive stick-breaking weights
%
\begin{equation}
\label{stick-breaking-WF} w_{1}(t)=v_{1}(t),\qquad w_{i}(t)=v_{i}(t)
\prod_{j<i}\bigl(1-v_{j}(t)\bigr),\qquad
v_{i}(\cdot)\stackrel{\mathrm{i.i.d.}} {\sim} \operatorname{WF}(1,\theta).
\end{equation}
The process $w(\cdot)=\{w(t),t\ge0\}$ defined above for the vector of
weights $w(t)=(w_{1}(t),  w_{2}(t),\ldots)$ has been characterised by
Feng and Wang \cite{FW07}, {who investigate its sample path properties.}
It is clear that the constraints $a=1$ and $b=\theta$ are by no means
essential but only chosen to preserve the connection with the Dirichlet
process (see Proposition~\ref{thmdep-RPM} below). Section~\ref{secdiscussion} will briefly discuss possible extensions.
With this formulation, (\ref{dependentprocess}) defines a family of
dependent random probability measures which retain at every time point
the stick-breaking structure featured by (\ref{stick-breaking-weights}).
For ease of reference, we summarise the construction in the following
definition.

\begin{definition}\label{diffusiveDP}
A family of dependent random probability measures $P=\{P_{t}, t\ge0\}$
with representation (\ref{dependentprocess})--(\ref
{stick-breaking-WF}) is said to be a \emph{diffusive Dirichlet process}.
\end{definition}

Besides studying $w(\cdot)$, Feng and Wang \cite{FW07} also consider a
construction
more general than~\eqref{dependentprocess}, where the atoms
$(x_{1},x_{2},\ldots)$ are let to be a Markov process on $\X^{\infty
}$. However, their model is too general for our purposes and its
properties are hard to establish without further assumptions on the
model components. Hence, we need to formalise the path properties of
$P$, as these cannot be deduced directly from either of the models
considered in their paper.

To this end, denote by $C_{\P(\X)}([0,\infty))$ the space of continuous
functions from $[0,\infty)$ to $\P(\X)$. Here $P$ denotes the $C_{\P
(\X
)}([0,\infty))$-valued random element $\{P_{t},t\ge0\}$, and $P_{t}$
its coordinate projection at $t$, so that $P_{t}\in\P(\X)$. We endow
$\P
(\X)$ with the topology induced by the total variation norm, so that
elements of $C_{\P(\X)}([0,\infty))$ have modulus of continuity
\[
\omega(P,\delta)=\sup_{|s-t|<\delta}\sup_{A \in\B(\X
)}\bigl|P_{t}(A)-P_{s}(A)\bigr|,
\]
and $P\in C_{\P(\X)}([0,\infty))$ if and only if $\omega(P,\delta
)\to0$
as $\delta\to0$, that is if $P_{t+s}\to P_{t}$ in total variation
distance as $s\to0$. See Billingsley \cite{B68}, Chapter~2.
The fact that the diffusive Dirichlet Process $P$ in Definition~\ref
{diffusiveDP} {has continuous sample paths in total variation}, should
be intuitive from the construction, since the only time-varying
quantities are diffusion processes mapped through a continuous
function. The following proposition formalises this fact.

\begin{proposition}\label{propositiondiffusion}
Let $P$ be as in Definition~\ref{diffusiveDP}. Then $P$ is a Feller
process with realisations almost surely in $C_{\P(\X)}([0,\infty))$.
\end{proposition}

The Feller property for $P$ guarantees certain desirable path
properties which, among other things, yield the well-definedness of the
process and its Markovianity. We refer the reader to Ethier and Kurtz
\cite{EK86},
Chapter~4, for more details on Feller operators. However, the
continuity of sample paths is not implied by the Feller property and is
proven separately. In particular, it is such continuity that will
allow, after embedding $P$ in an appropriate statistical model, to
select almost surely continuous functions as in Assumption~\ref{assumption}.

Since the stationary distribution of the Wright--Fisher diffusion, used
in \eqref{stick-breaking-WF}, is the same distribution used in (\ref
{stick-breaking-weights}) for the stick-breaking components, it is also
intuitive that the process of Definition~\ref{diffusiveDP} is
stationary with respect to the law of a Dirichlet process, as stated by
the next proposition.

\begin{proposition}\label{thmdep-RPM}
Let $P$ be as in Definition~\ref{diffusiveDP} and, for a finite
nonnull measure $\alpha$ on $\X$, let $\D_{\alpha}$ denote the law of
a Dirichlet process. {Then $P$ is reversible and stationary with
respect to $\D_{\alpha}$. In particular, if} $P_{0}\sim\D_{\alpha}$,
then $P_{t}\sim\D_{\alpha}$ for every $t>0$.
\end{proposition}

Hence, the marginal states of the diffusive Dirichlet process are
Dirichlet distributed.
Here it is important to note that modelling the above measure-valued
process as stationary will not constrain the data to come from a
stationary process. This will be more transparent when we will consider
the hierarchical statistical model for the data. On the contrary, this
aspect will turn into an advantage since it will allow to propagate in
time the support properties of the Dirichlet prior. Indeed it is well
known that the Dirichlet prior has full weak support. That is, if $\X$
is the support of the parameter measure $\alpha$, then the support of
$\D_{\alpha}$ in the weak topology is
%
\begin{equation}
\label{DP-support} \operatorname{supp}(\D_{\alpha})=\bigl\{Q\in\P(\X)\dvt
\operatorname {supp}(Q)\subset\X\bigr\}.
\end{equation}
See Ghosh and Ramamoorthi \cite{GR03}, Section~3.2.3. Proposition~\ref
{thmdep-RPM} then
implies that at stationarity $P_{t}$, marginally, has support (\ref
{DP-support}). See also Barrientos, Jara and Quintana \cite{BJQ12} for
sufficient conditions for
having full weak support in the context of dependent stick-breaking processes.

Another byproduct of Proposition~\ref{thmdep-RPM} is the immediate
derivation of the marginal moments of~$P$. In particular, let
$P_{0}\sim\D_{\alpha}$, with $\alpha=\theta G$, where $\theta>0$ and
$G\in\P(\X)$. Then, for all $t\ge0$ and $A\in\B(\X)$,
%
\begin{equation}
\label{DP-moment} \E\bigl[P_{t}(A)\bigr]=G(A),\qquad \operatorname{Var}
\bigl[P_{t}(A)\bigr]=\frac{G(A)(1-G(A))}{\theta+1},
\end{equation}
where $P_{t}(A)$ denotes \eqref{dependentprocess} evaluated at the
set $A$.
In addition, the following proposition provides an explicit expression
for the autocorrelation function of the process.

\begin{proposition}\label{propacf}
Let $P$ be as in Definition~\ref{diffusiveDP}. Then, for any $A\in\B
(\X
)$ and any $t,s>0$,
{
%
\begin{equation}
\label{acf} \operatorname{Corr}\bigl(P_{t}(A),P_{t+s}(A)\bigr)=
\frac{(1+\theta)[(2+\theta)+\theta
\e^{-\lambda s}]}{
(2+\theta)(1+2\theta)-\theta \e^{-\lambda s}},
\end{equation}
where $\lambda=(1+\theta)/2$.}
\end{proposition}

As expected, the correlation does not depend on the set $A$, since the
time dependence enters only via the weights and not the locations.
Furthermore, it is easily seen that the correlation decays
exponentially to $(1+\theta)/(1+2\theta)$ as $s\rightarrow\infty$.
{Although this can perhaps be considered as an undesirable property,}
the existence of a lower bound for the correlation is a common feature
of all dependent processes whose atoms are fixed (see Rodriguez and
Dunson \cite{RD11}).
{In Section~\ref{secdiscussion}, we will provide more comments on
this point and outline a possible extension which aims at relaxing some
of the model constraints.}

The above dependent process can be used to formulate a dependent
mixture model by considering the time-varying density
%
\begin{equation}
\label{depmixturemodel} f_{P_{t}}(y)=\int K(y|x)P_{t}(\mathrm{ d}x),
\end{equation}
where $K(\cdot| x)$ is a kernel density with parameter $x$. An
equivalent formulation is provided in the form of the hierarchical model
%
\begin{eqnarray}
\label{hierarchydiff-DP} %
y|x &\sim& K(y|x),
\\
x|t,P_{t} & \sim& P_{t},\nonumber
\\
P & \sim& \operatorname{diff\mbox{-}DP}, \nonumber
\end{eqnarray}
where $P\sim\operatorname{diff\mbox{-}DP}$ denotes that $P$ is a diffusive
Dirichlet process.
Since $P$ is the only time-varying component in the statistical model,
the dependent mixture inherits the diffusive behavior from $P$, and
Proposition~\ref{thmdep-RPM} implies that marginally $f_{P_{t}}$ is a
Dirichlet process mixture.
Thus, the dependent mixture induces a prior distribution on the space
$C_{\P(\X)}([0,\infty))$ of continuous functions from $[0,\infty)$ to
$\P(\X)$, which almost surely selects functions {$g\dvtx \X\times
[0,T]\rightarrow\R_{+}$} that satisfy Assumption~\ref{assumption}. The
choice of kernel $K$ determines the discrete or continuous nature of
the sections $g(\cdot,t)=f_{P_{t}}(\cdot)$ and their support.


\section{Posterior computation}\label{postcomputation}

We overview the strategy for simulating from the posterior distribution
of \eqref{depmixturemodel} and of other quantities of interest, such
as, for example, the mean functional process $\eta_t=\bb
{E}_{f_{P_t}}(y)$, which depicts the average stochastic process driving
the observations. Here we highlight the main points of interest, while
the fully detailed procedure can be found in Appendix~\ref{appa}.
For notational simplicity, and in view of the real data example below,
we assume a single data setting. However, the strategy allows a
straightforward extension to the case of multiple observations at every
time point.

Specifically, we assume data ${y}^{(n)}=(y_{t_1},\ldots,y_{t_n})$ are
observed at times $0\le t_1< \cdots< t_n$, where time intervals are not
necessarily equally spaced.
The target of inference is the data generating time-varying
distribution $g(y,t)$, such that $y_{t_{i}}\sim g(\cdot,t_{i})$. We
model such $g$ by means of a diffusive Dirichlet mixture, with $g(\cdot
,t)=f_{P_{t}}(\cdot)$. To this end, let
$v(\cdot)=((v_{1}(t),v_{2}(t),\ldots), t\ge0)$ denote a collection of
independent Wright--Fisher diffusions defined as in (\ref{wf-sde}),
with $v_j(\cdot)\sim^{\mathrm{i.i.d.}} \operatorname{WF}(1,\theta)$. Let also
$x=(x_{1},x_{2},\ldots)$ be the random locations sampled from a
nonatomic probability measure $G$.
Hence, the data generating process is modelled as
%
\begin{equation}
\label{dnpm} {f}_{P_t}\bigl(y\mid v(t), \msf{x} \bigr)=\sum
_{j\geq1} \msf{w}_j(t) K(y\mid x_j ),
\end{equation}
with $\msf{w}_j(t)$ as in (\ref{stick-breaking-WF}).
The model induces dependence among the observations, which are
exchangeable for fixed $t$ but partially exchangeable in general.

The infinite dimensionality of the random measure at $t$ is dealt with
via slice sampling (Damien, Wakefield and Walker \cite{DWW99}, Walker
\cite{W07}). Specifically, we extend the slice
algorithm in Kalli, Griffin and Walker \cite{KGW11} to augment the
above random density by
%
\begin{equation}
\label{dnpmaug2} {f}_{P_t}\bigl(y,u,s\mid\msf{v}(t), \msf{x} \bigr)=
\frac{\bb{I}(u< \psi
_{s})}{\psi_{s}} \msf{w}_s(t) \msf{K}(y\mid{x}_s ),
\end{equation}
where $s\mapsto\psi_s$ is a decreasing function with known inverse
$\psi^*$, for example, $\psi_s=\e^{-\eta s}$, for $0\leq\eta\leq1$. The
latent variable $s$ indexes {which of the kernels $K(\cdot\mid x_{s})$
better captures the mass at $y$, and given $s$, $\msf{u}\sim\msf
{U}(0,\psi_s)$}. For the purpose of estimation, it is enough to
condition on $v^{(n)}=  \{v(t_i)  \}_{i=1}^n$, rather than
on the
whole path $v(\cdot)$. In this section and in the \hyperref[app]{Appendix}, we will use
the notation $\mathcal{L}(z)$ to indicate generically the law of {$z$}.
The conditional augmented likelihood is given by
%
\begin{equation}
\mathcal{L} \bigl({y}^{(n)}, {u}^{(n)}, {s}^{(n)}
\mid v^{(n)},x \bigr) = \prod_{i=1}^n
{\frac{\bb{I}(u_i< \psi_{s_i})}{\psi_{s_i}}} \biggl[v_{s_i}(t_i)\prod
_{k<s_i}\bigl(1-v_k(t_i)\bigr) \biggr]
K(y_{t_i}\mid{x}_{s_i}),
\end{equation}
where $\msf{u}^{(n)}=(u_1,\ldots,u_n)$ and $\msf
{s}^{(n)}=(s_1,\ldots
,s_n)$, with $u_i={u_{t_i}}$ and $s_i={s_{t_i}}$. Due to the random
truncation induced by the slice sampling method, one is only able to
learn about the first $m$ Wright--Fisher processes and locations, where
%
\begin{equation}
\label{m=max} \msf{m}=\max\bigl(\bigl\lfloor\psi^*(u_{1})\bigr
\rfloor,\bigl\lfloor\psi ^*(u_{2})\bigr\rfloor ,\ldots,\bigl\lfloor
\psi^*(u_{n})\bigr\rfloor\bigr),
\end{equation}
and $\lfloor A\rfloor$ denotes the integer part of $A$.
Hence, denoting
%
\begin{equation}
\label{1mnotation} {v}_\msf{1:m} ^{(n)}=\bigl({v}_1
^{(n)},\ldots,{v}_m ^{(n)}\bigr),\qquad
x_{1:m}=(x_1,\ldots,x_m),
\end{equation}
we see that
%
\[
\mathcal{L} \bigl(\msf{v}_\msf{1:m} ^{(n)},
\msf{x}_\msf{1:m}\mid{y}^{(n)} ,\msf {u}^{(n)},
\msf{s}^{{(n)}} \bigr) \propto\mathcal{L} \bigl({y}^{(n)},
\msf{u}^{(n)}, \msf{s}^{{(n)}} \mid \msf {v}_\msf{1:m}
^{(n)},\msf{x}_\msf{1:m} \bigr) \mathcal{L}\bigl(\msf
{v}_\msf {1:m} ^{(n)}\bigr) \mathcal{L}(\msf{x}_\msf{1:m})
,
\]
while the posterior distribution remains the unchanged prior for all
$\msf{v}_l ^{(n)}$,
$l>m$.
Here the $\msf{m}$ processes and locations are {mutually} independent,
so that
\[
\mathcal{L}\bigl(\msf{v}_\msf{1:m} ^{(n)}\bigr)=\prod
_{j=1}^{\msf
{m}}\mathcal {L}\bigl(
\msf{v}_j ^{(n)}\bigr),\qquad \mathcal{L}(x_{1:m})=\prod
_{j=1}^{\msf{m}}\mathcal{L}(\msf{x}_j
).
\]
It remains to observe that the finite dimensional distributions for
each Wright--Fisher process are
%
\begin{equation}
\label{lawv} \mathcal{L}\bigl(\msf{v}_j ^{(n)}\bigr)=
\pi_{\msf{v}}\bigl(\msf{v}_j(t_1)\bigr)\prod
_{i=2}^n\msf{p}_{\msf{v}}\bigl(
\msf{v}_j(t_i)\mid\msf{v}_j(t_{i-1})
\bigr),
\end{equation}
where $\pi_\msf{v}=\operatorname{Beta}(1,\theta)$ and $\msf{p}_{\msf{v}}$
denotes the transition density of the Wright--Fisher diffusion. This is
known {explicitly} (Ethier and Griffiths \cite{EG93}), but has an
infinite series
representation. In view of a further implementation of the slice
sampler, we resort to the representation of $\msf{p}_{\msf{v}}$ due to
Mena and Walker \cite{MW09}, which reads
%
\begin{equation}
\label{WFtransition} \msf{p}_{\msf{v}}\bigl(\msf{v}(t)\mid\msf{v}(0)\bigr)=\sum
_{m=0}^\infty r_{t}(m) D\bigl(
\msf{v}(t)\mid m,\msf{v}(0)\bigr),
\end{equation}
where $r_{t}(m)$ is an appropriate deterministic function and $D(\msf
{v}(t)\mid m,\msf{v}(0))$ a finite mixture of Beta distributions. This
leads to an augmentation of $\msf{p}_{\msf{v}}$ similar to that
outlined above for $f_{P_{t}}$, allowing to avoid the infinite computation.

\begin{algorithm}[t!]
\begin{algorithmic}[1]
\footnotesize
\STATE\textbf{input:} data ${(y_{t_1},\ldots,y_{t_n})}$ and their
recording times $(t_1,\ldots,t_n)$
\STATE\textbf{initial values for:} \\
\sf{(a)} Hyper-parameters in $G$ and WF parameters\\
\sf{(b)} Upper limit $\msf{m}^{(0)}$ for the number of locations and WF
processes\\
\sf{(c)} Membership variables $\msf{s}^{(0)}=(s_{t_1},\ldots,s_{t_n})$,
taking values in $\{1,\ldots,\msf{m}^{(0)}\}$ \\
\sf{(d) }WF processes at $(t_1,\ldots,t_n)$ and corresponding latent
variables\\ 
\sf{(e)} Location values 
\FOR{${ I}=1$ to $\msf{\mathit{ITER}}$}
\STATE Sample slice variables $\msf{u_{t_i}}\sim\msf{U}(0,\psi
_{s_{t_i}})$ and update the value of $\msf{m}_{I}$
\IF{$\msf{m}^{( I)}>\msf{m}^{({ I}-1)}$}
\STATE Take extra WF processes and their latent variables 
from priors
\ENDIF
\STATE Update transition density latent variables 
\STATE Update WF processes values
\STATE Update location values
\STATE Update WF parameter values
\STATE Update membership variables $\msf{s}^{({ I})}$ taking
values in $\{1,\ldots,\msf{m}^{({ I})}\}$
\ENDFOR
\end{algorithmic}
\caption{Gibbs sampler for diffusive DP mixtures}\vspace*{6pt}\label{algDiffDP}
\end{algorithm}
%

With the above specification, a Gibbs sampler algorithm can be designed
as in Algorithm \ref{algDiffDP}.
The reader is referred to Appendix~\ref{appa} for the algorithm
details not included in this section.


\section{Illustration}\label{secillustration}
Following the above framework, in this section we illustrate an
application of the diffusive Dirichlet process with simulated and real
data. More specifically, we consider:
\begin{longlist}[(ii)]
\item[(i)] simulated observations sampled at equally spaced intervals
from the time-dependent normal density
%
\begin{equation}
\label{toy} \mathrm{N}\bigl(\operatorname{cos}(2t)+t/2,1/10\bigr);
\end{equation}
\item[(ii)] 300 real observations given by daily exchange rates between
US dollars and Mexican peso during the period from September 26th,
2008, to December 7th, 2009.
\end{longlist}

The assumption of data collected at regular intervals here is for
computational simplicity only, as the model, through \eqref{WFtransition}, allows for not equally spaced samples.

In order to complete the specification of the diffusive mixture, we use
a conjugate vanilla choice for the kernel $K$ and centering measure
$G$. Specifically, let
\begin{eqnarray*}
K(y\mid x)&=&\mathrm{N}\bigl(y\mid m,v^{-1}\bigr),
\\
G(x)&=&\mathrm{N} \bigl(\msf{m}\mid0,1000v^{-1} \bigr)\operatorname
{Gamma} (v\mid10,1 ),
\end{eqnarray*}
with $x=(m,v)$. The parametrization for $G$ is chosen to achieve a
large variance at the location level, to cover all observations with
high probability. This is required, as the locations are random but
fixed over time, and the weight processes should be able to pick any
good candidate location within the data state space at a given time.
Running Algorithm \ref{algDiffDP} allows to draw posterior inferences
for any functional of the diffusive Dirichlet mixture model. In
particular, besides the time-varying density, we are interested in the
mean functional
%
\begin{equation}
\label{meanfunc} \eta_{t}=\int_{\bb{R}} y
f_{P_{t}}(\mathrm{d}y).
\end{equation}
All examples {in this section} are based on 2000 effective iterations
drawn from 10,000 iterations thinned each five and after a burn in
period of 5000 iterations. We verified practical convergence using the
convergence diagnostics of Gelman and Rubin \cite{GR92} and of Raftery
and Lewis \cite{RL92}, and neither
showed any evidence of convergence problems.\looseness=1

\begin{figure}

\includegraphics{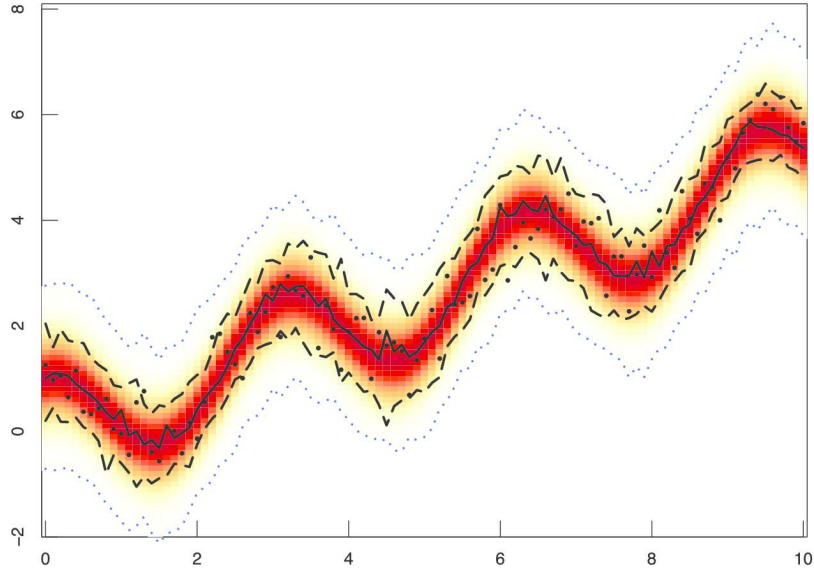}

\caption{MCMC-based estimation for single data points
(dots) sampled at equally spaced intervals from \protect\eqref
{toy}. The
picture shows the true model (heat contour), the pointwise posterior
mode (solid line) and 95\% credible intervals for the mean
functional (dotted lines), and the pointwise 95\% quantiles of
the posterior estimate of the time-varying density (dotted
lines).}\vspace*{4pt}\label{figtoy}
\end{figure}

Figure~\ref{figtoy} shows the results corresponding to a first dataset
simulated from \eqref{toy}, where 100 single observations are collected
at equally spaced intervals. The true data generating process is shown
as a heat contour,
with darker regions being those of higher probability, and presents traits of
nonstationarity and seasonality. The dots are the simulated data, the solid
and dashed lines are the pointwise posterior mode and 95\% credible intervals
for the mean functional, respectively, and the dotted lines are
the pointwise 95\% quantiles of the posterior estimate of the
time-varying density. The picture shows that the model correctly
captures the nonstationary behaviour with a strong trend and
seasonalities. Furthermore, even in this setting with structural lack
of instant-wise information, due to the availability of only single
data points, the uncertainty of the estimates is reasonably low, as the
model instantaneous variance is indirectly learned by the algorithm
from the overall data pool. The smoothness level captured by the
estimation is also acceptable, given the above considerations on the
lack of information and considered that the model is ultimately based
on Wright--Fisher components.

\begin{figure}

\includegraphics{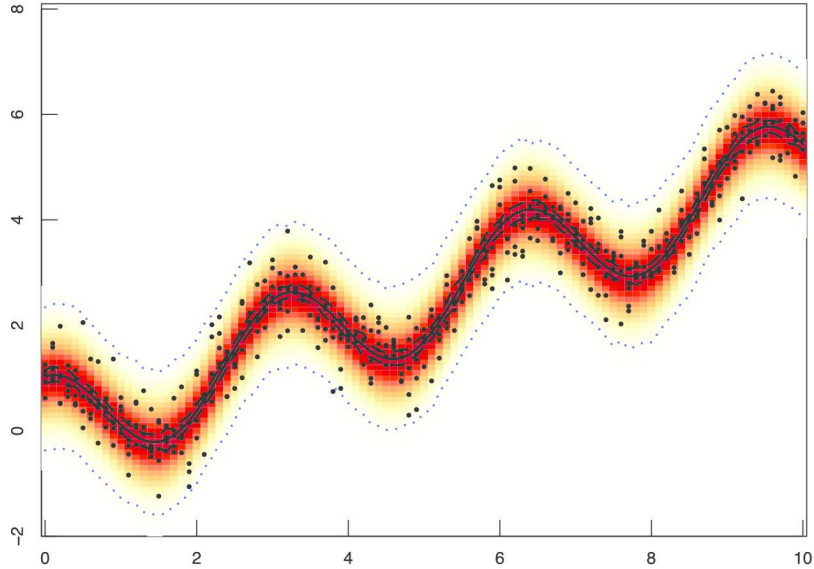}

\caption{MCMC-based estimation for multiple data points
(dots) sampled at equally spaced intervals from~\protect\eqref
{toy}. The
picture shows the true model (heat contour), the pointwise posterior
mode (solid line) and 95\% credible intervals for the mean
functional (dotted lines), and the pointwise 95\% quantiles of
the posterior estimate of the time-varying density (dotted
lines).}\label{figtoyMulti}
\end{figure}

In order to investigate the degree of improvement, one can gain with
multiple data points, we performed the same type of inference as in
Figure~\ref{figtoy} on a second set of data\vadjust{\eject} simulated from \eqref{toy},
where at each of the 100 time points, five data are available. The
results, reported in Figure~\ref{figtoyMulti}, show that the accuracy
of the estimation increases satisfactorily, as the true model behaviour
is captured with considerably less uncertainty. This is especially true
for the mean functional (solid line), whose credible intervals
(dashed lines) are very narrow. It is also important to note that
the smoothness of the true model is also correctly learned by the
estimates. This is due to the fact that the rough behaviours of the
model subcomponents are confined to levels of the hierarchy where their
impact on the final estimation, in presence of enough information, is
pooled together and softened by adapting the component-specific
volatilities. This however does not prevent the model from capturing
quick deviations from a smooth trend, as showed by the next example.\looseness=1

For the illustration with real observations, we concentrate on a
challenging set of single data points per observation time, which again
provides scarce instantaneous information. Financial data sets as in
(ii) are often described with parametric state-space models. This can
put serious constraints on the ability of the model to capture the
correct marginal distributions and quick deviations from the general
pattern. By making the nonparametric assumption that the state-space
model follows a diffusive Dirichlet process mixture, we are relaxing
such constraints and granting great flexibility to the model. In the\vadjust{\eject}
interpretation related to hidden Markov models, the unobserved signal
here models an evolving distribution, driven by a measure-valued Markov
process, and the observations are sampled from the signal states. Note
however that this framework does not impose any Markovianity nor
stationarity on the observations.

\begin{figure}

\includegraphics{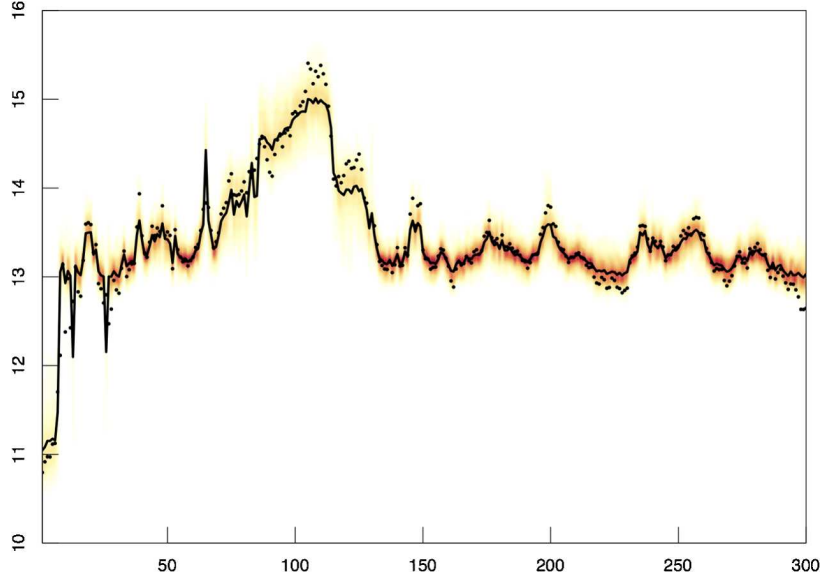}

\caption{MCMC-based pointwise posterior density estimate
(heat contour) and {pointwise posterior mode of the mean functional
(solid line)}, based on single data points (dots) corresponding
to dataset (ii).}\label{pltfin}
\end{figure}

\begin{figure}

\includegraphics{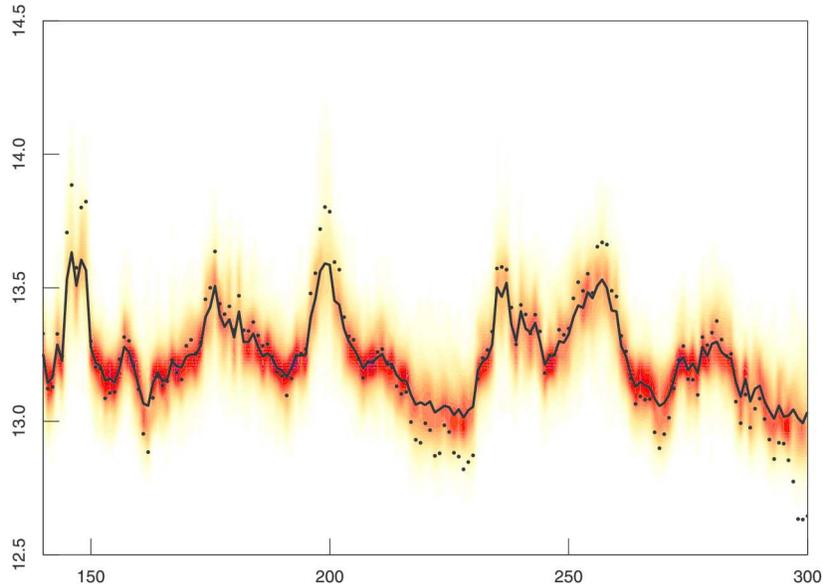}

\caption{A sub-region of Figure \protect\ref{pltfin}.}\label{pltfin2}
\end{figure}

Figure~\ref{pltfin} shows the results on the exchange rate data set,
with the horizontal axis representing time and the vertical axis
representing the index value.
The heat contour outlines the shape of the pointwise posterior estimate for
the time-varying density function $f_{P_{t}}$, with darker regions corresponding to
higher posterior probability. The solid line is
{the pointwise
mode of the posterior} mean functional \eqref{meanfunc}.
The model is able to capture highly volatile behaviours, such as that
encountered in the period between 70 and 130. These kind of changes are
typically not well recognised by {parametric models for time series},
which are too rigid to allow for unexpected detours. Figure~\ref{pltfin2} shows a sub-region of Figure~\ref{pltfin}, where sudden
spikes and different local trends are also shown to be correctly
captured, regardless of how abrupt these appear.
Another important aspect to be noted is the fact the regions of high
estimated probability need not correspond to the regions where data are
observed, even having single data points, as in the central sub period
around day $225$ in Figure~\ref{pltfin2}. This feature is determined by
the model dependence, which allows to borrow information across time,
and should not be confused with model rigidity, as the model clearly
captures sudden deviations from the trend as that occurring around time 200.

\section{Discussion, extensions and future work}\label{secdiscussion}

We introduced a new class of prior distributions on the space of
time-indexed, $t$-continuous functions $g\dvtx \X\times[0,T]\rightarrow\R
_{+}$, such that $g(\cdot,t)$ is a density for all $t\in[0,T]$. Such
priors are induced by diffusive Dirichlet process mixtures, which
extend the Dirichlet process mixture model of Lo~\cite{L84} to a framework
of Feller measure-valued processes with continuous trajectories.
The resulting dependent random density \eqref{depmixturemodel} can be
used to tackle various statistical problems of interest in many fields
such as econometrics, finance and medicine among others, when the
random phenomena evolving in continuous time are not satisfactorily
modelled parametrically. On the other hand, it can be an alternative to
other nonparametric dependent models which are statistically
intractable. For example, when the underlying model structure is linked
to population dynamics, it could be desirable to model data by means of
Fleming--Viot processes (see Ethier and Kurtz \cite{EK93}). However,
such a process
is computationally intractable in view of inference, and the presented
model can then be used to this end, as the main properties of the
former such as the stationary measure and path regularity are preserved.

{Overall, the introduced model exhibits a good mix of flexibility and
structure, yet leaving room for computational efficiency. The full
support property of the Dirichlet process and the time dependence of
Wright--Fisher diffusions are combined to yield an equilibrated
compromise between adaptivity and dependence, and the Gibbs sampler
with slice steps is simple to implement. These aspects lead to the
ability of jointly detecting smooth and irregular evolutions of the
distribution which generates the observations, while borrowing strength
between observations when this is needed.}

We briefly outline two possible extensions of the model, concerned with
certain features that can nonetheless be considered as constraints in
certain contexts. The first is a reparametrisation which retains the
overall structure, and aims at broadening the set of stationary
distributions of the model. The second changes the qualitative features
of the model and aims at removing some rigidities. These are concerned
with the fact that the correlation is bounded from below, together with
the fact that the dependence structure imposes some restrictions on the
data generating mechanism.

The construction can be easily extended by relaxing the assumption of
identity in distribution of the WF diffusions used in \eqref
{stick-breaking-WF}. This can be done, for example, by considering the
class of GEM diffusions, also developed in Feng and Wang \cite{FW07},
to allow for
more general stationary distributions of the weights, such as
two-parameter Poisson--Dirichlet distributions (Pitman \cite{P95},
Pitman and Yor \cite{PY97}) or GEM
distributions (Johnson, Kotz and Balakrishnan \cite{JKB97}, Chapter~41). A general construction of
the latter class of random measures is obtained by taking $v_{i}\sim
^{\mathrm{ind}} \operatorname{Beta}(a_{i},b_{i})$ in \eqref{stick-breaking-weights},
while the former corresponds to choosing $a_{i}=1-\sigma$ and
$b_{i}=\theta+i\sigma$ for all $i\ge1$.
By analogy with the construction in Section~\ref{secDPM}, GEM-type
measure-valued diffusions can be defined by letting $v_{i}(\cdot)\sim
^{\mathrm{ind}}\operatorname{WF}(a_{i},b_{i})$ in \eqref{stick-breaking-WF},
that is the
collection of Wright--Fisher diffusions which induce the
time-dependence in the weights of (\ref{dependentprocess}) is given by
independent, and no longer identically distributed, processes. An
appropriate extension of Proposition~\ref{thmdep-RPM} easily follows.
One can then replace $P$ in (\ref{hierarchydiff-DP}) with the
resulting measure-valued GEM process to yield additional modelling
flexibility. The resulting class of dependent mixture models extends,
to a time-dependent framework, the priors considered in Ishwaran and
James \cite{IJ01}.
Of course there is a trade-off between the amount of flexibility one
pursues and the amount of parameters one is willing to deal with in
terms of computational effort. The choice of two-parameter
Poisson--Dirichlet distributions probably guarantees extra flexibility
at almost no extra cost, as it allows to control more effectively the
posterior distribution of the number of clusters (Lijoi, Mena and Pr\"
unster \cite{LMP07}), and
only requires one additional step at every Gibbs sampler iteration for
updating the posterior distribution of $\sigma$.

A different direction can be considered with the aim of relaxing the
dependence structure the model imposes on the data. This is the object
of a currently ongoing work by the authors, of which we concisely
outline the main ideas.
We consider a specific choice for the general dependent process
\[
P_{t}^{(\gamma)}=\sum_{i=1}^{\infty}w_{i}(t)
\delta_{x_{i}(t)},\qquad t\ge0,
\]
parametrised by $\gamma>0$, which keeps the model complexity and the
implied computational burden relatively low, but yields an
autocorrelation which vanishes exponentially fast and whose structure
can be inferred from the data. This is obtained, for example, by
letting the weights $w(\cdot)$ be as in \eqref{stick-breaking-WF}, and
by letting the initial atoms $x_{i}(0)\sim^{\mathrm{i.i.d.}}G$ be updated one at a
time after an interval with $\operatorname{Exp}(\gamma)$ distribution from the last
update. The atom to be replaced is chosen, for example, according to a fixed
distribution, and the update is chosen from $G$. The notation
$P^{(\gamma)}$ for the resulting model highlights the role of the
intensity $\gamma$ of the underlying Poisson point process which
regulates how often these innovations occur.
Such specification extends the model of Section~\ref{secDPM} to a
model which is still stationary with respect to the law of a Dirichlet
process, and has a correlation which decreases to zero exponentially
with speed regulated by $\gamma$.
Informally, $P^{(\gamma)}$ interpolates between the diffusive Dirichlet
process, obtained as an appropriate limit of $P^{(\gamma)}$ as $\gamma
\to0$, and a purely discontinuous model $P^{(\infty)}$, whereby in
every finite interval infinitely many atoms are updated, so that for
every $s>0$, $P^{(\infty)}_{t}$ and $P^{(\infty)}_{t+s}$ are
uncorrelated. This adds great flexibility to the dependence structure,
which can be learned from the data by implementing $P^{(\gamma)}$ in a
hierarchical model, similar to \eqref{hierarchydiff-DP}, augmented
with a prior distribution on $\gamma$.
However, besides the clear advantages related to the correlation which
is no longer constrained, the model sample paths are no longer
continuous but only continuous in probability, and the Markov property
is retained only with respect to the filtration generated by $(w(\cdot
),x(\cdot))$ and not with respect to the natural filtration.
A further alternative, which retains the path continuity, is the
possibility of allowing also the atoms to diffuse. {However,} this way
seems difficult if one is interested in {proving} minimal theoretical
properties for the model.

\begin{appendix}\label{app}


\section{Proofs}

\subsection*{Proof of Proposition \protect\ref
{propositiondiffusion}}

{Define
\[
\Delta_{\infty}= \biggl\{z\in[0,1]^{\infty}\dvt \sum
_{i\ge
1}z_{i}=1 \biggr\},
\]
and let, for any fixed sequence $x=(x_{1},x_{2},\ldots)\in\X^{\infty}$,
$\P_{x}$ be the set of purely atomic probability measures with support
$x\in\X^{\infty}$. Denote by $\varphi_{x}\dvtx \Delta_{\infty
}\rightarrow\P
_{x}$ the transformation $\varphi_{x}(\phi(v))=\sum_{i=1}^{\infty
}w_{i}\delta_{x_{i}}$,}
where $v=(v_{1},v_{2},\ldots)$ and $\phi\dvtx [0,1]^{\mathbb
{N}}\rightarrow
\Delta_{\infty}$ is defined as
\[
\phi_{1}(v)=v_{1},\qquad \phi_{i}(v)=v_{i}(1-v_{1})
\cdots(1-v_{i-1}),\qquad i>1.
\]
Note that the map $\phi$ is a bijection, with $v_{i}=w_{i}/(1-\sum_{k=1}^{i-1}w_{k})$,
and that $\varphi_{x}$ is continuous as
a function of $w=\phi(v)$ in total variation norm, since for every
$\varphi_{x}(w)$ and $\varepsilon>0$ we can find a neighbourhood
%
\begin{equation}
\label{neigh} U(w,\varepsilon)= \biggl\{w'\in\Delta_{\infty}
\dvt \sum_{i\ge
1}\bigl|w_{i}-w'_{i}\bigr|<
\varepsilon \biggr\}
\end{equation}
so that $w^{*}\in U(w,\varepsilon)$ implies $\varphi_{x}(w^{*})\in
U_{\mathrm{TV}}(\varphi_{x}(w),\varepsilon)$, with
\begin{equation}
\label{TVneighbourhood} U_{\mathrm{TV}}\bigl(\varphi_{x}(w),\varepsilon
\bigr)= \bigl\{\varphi _{x}\bigl(w'\bigr)\in\P
_{x}\dvt d_{\mathrm{TV}}\bigl(\varphi_{x}(w),
\varphi_{x}\bigl(w'\bigr)\bigr)<\varepsilon \bigr\},
\end{equation}
since
\[
d_{\mathrm{TV}}\bigl(\varphi_{x}(w),
\varphi_{x}\bigl(w'\bigr)\bigr)=\sup_{A\in\B(\X
)}
\biggl|\sum_{i\ge1}w_{i}\delta_{x_{i}}(A)-
\sum_{i\ge1}w'_{i}\delta
_{x_{i}}(A) \biggr| \le \sum_{i\ge1}\bigl|w_{i}-w'_{i}\bigr|
< \varepsilon.
\]
Furthermore, $\varphi_{x}(w)$ is invertible in $w$, with continuous inverse
%
\begin{equation}
\label{psi-inverse} \varphi_{x}^{-1}(P)=\bigl(P\bigl(
\{x_{1}\}\bigr),P\bigl(\{x_{2}\}\bigr),\ldots\bigr)=w.
\end{equation}
%
This implies that we can define a Feller semigroup $\{T_{x}(t)\}_{t\ge
0}$ on $C(\P_{x})$ by means of $T_{x}(t)\psi=[S(t)(\psi\circ\varphi
_{x})]\circ\varphi_{x}^{-1}$,
where $\{S(t)\}_{t\ge0}$ is the Feller semigroup on
$C(\Delta_{\infty})$ corresponding to the process $w(\cdot)$ and
$\varphi_{x}^{-1}$ is as in (\ref{psi-inverse}).
Theorem~4.2.7 of
Ethier and Kurtz \cite{EK86} now implies that for every probability
measure $\nu$ on $\P
_{x}$, there exists a Markov process $P$ corresponding to $\{T_{x}(t)\}
_{t\ge0}$ with initial distribution $\nu$ and sample paths in $D_{\P
_{x}}([0,\infty))$, the space of right-continuous functions from
$[0,\infty)$ to $\P_{x}$ with left limits, {equipped with the Skorohod
topology. See Billingsley~\cite{B68}, Chapter~3, for details. Moreover,
being a
continuous bijection with continuous inverse, for any fixed $x$,
$\varphi_{x}$ is a homeomorphism of $\Delta_{\infty}$ into $\P_{x}$,
from which $\P_{x}$ is locally compact and separable.
Denote now with $p_{1}(t,P,\mathrm{ d}P')$ and $p_{2}(t,w,\mathrm{ d}w')$ the
transition functions corresponding to the semigroups $\{T_{x}(t)\}
_{t\ge
0}$ and $\{S(t)\}_{t\ge0}$, respectively, and define $U(w,\varepsilon)$
as in \eqref{neigh} and $U_{\mathrm{TV}}(P,\varepsilon)$ as in
\eqref{TVneighbourhood}. Then for every $P\in\P_{x}$ and $\varepsilon>0$, we have
%
\begin{equation}
\label{no-jumps-condition} t^{-1}p_{1}\bigl(t,P,U_{\mathrm{TV}}(P,
\varepsilon )^{c}\bigr)=t^{-1}p_{2}\bigl(t,w,U(w,
\varepsilon)^{c}\bigr)\rightarrow0 \qquad\mbox{as }t\rightarrow0,
\end{equation}
where the identity follows form the fact that the two events are
determined by the same subset of elementary events, and the right-hand
side of \eqref{no-jumps-condition} follows from the continuity of the
trajectories of $w(\cdot)$. The result now follows from Ethier and
Kurtz \cite{EK86},
Lemma~4.2.9.}

\subsection*{Proof of Proposition \protect\ref{thmdep-RPM}}%
{Since $w(\cdot)$ is reversible Feng and Wang \cite{FW07}, with each component
$v_{i}(\cdot)$ reversible with respect to a $\operatorname
{Beta}(1,\theta)$
distribution, and the atoms $x_{i}(t)\equiv x_{i}$ are trivially
reversible and independent of $w(\cdot)$, it follows that $P$ is
reversible. The full statement now follows by the fact that $x_{i}\sim
^{\mathrm{i.i.d.}}G$, and by assuming the initial distribution $v_{i}(0)\sim
\operatorname
{Beta}(1,\theta)$ for all $i\ge1$.}

\subsection*{Proof of Proposition \protect\ref{propacf}}
We have
\begin{eqnarray*}
\E\bigl(P_{t}(A) P_{t+s}(A)\bigr)&=&\E \biggl(\sum
_{i\ge1}w_{i}(t)\delta _{x_{i}}(A)\sum
_{j\ge1}w_{j}(t+s)\delta_{x_{j}}(A)
\biggr)
\\
&=& \E \biggl(\sum_{i\ge1}w_{i}(t)w_{i}(t+s)
\delta_{x_{i}}(A)
+\sum_{i\ge1}\sum_{j\ne i\ge1}w_{i}(t)w_{j}(t+s)
\delta _{x_{i}}(A)\delta_{x_{j}}(A) \biggr)
\\
&=& k_{s}G(A)+(1-k_{s})G^{2}(A),
\end{eqnarray*}
where
%
\begin{equation}
\label{k-s} k_{s} =\E \biggl(\sum_{i\ge1}w_{i}(t)w_{i}(t+s)
\biggr) =\sum_{i\ge1}\E\bigl(w_{i}(t)w_{i}(t+s)
\bigr).
\end{equation}
Here $k_{s}$ is independent of $t$ by stationarity
and $1-k_{s}$ is obtained by subtraction, since
\[
1=\sum_{i\ge1}w_{i}(t)\sum
_{j\ge1}w_{j}(t+s)= \sum
_{i\ge1}w_{i}(t)w_{i}(t+s)+\sum
_{i\ge1}\sum_{j\ne i\ge
1}w_{i}(t)w_{j}(t+s),
\]
from which
%
\begin{equation}
\label{covariance} \operatorname{Cov}\bigl(P_{t}(A),P_{t+s}(A)
\bigr)=k_{s}G(A) \bigl(1-G(A)\bigr),
\end{equation}
and, using \eqref{DP-moment},
%
\begin{equation}
\label{corrgamma0}
\nonumber
\mathrm{Corr}\bigl(P_{t}(A),P_{t+s}(A)
\bigr)=k_{s}(1+\theta).
\end{equation}
Now, from (\ref{stick-breaking-WF}) and using independence among the
$v_{i}(\cdot)$'s, we have
\begin{eqnarray*}
\E\bigl(w_{i}(t)w_{i}(t+s)\bigr) &=& \E \biggl[
\biggl(v_{i}(t)\prod_{j<i}
\bigl(1-v_{j}(t)\bigr) \biggr) \biggl(v_{i}(t+s)\prod
_{j<i}\bigl(1-v_{j}(t+s)\bigr) \biggr)
\biggr]
\\
&=& \E\bigl[v_{i}(t)v_{i}(t+s)\bigr]\prod
_{j<i}\E \bigl(\bigl(1-v_{j}(t)\bigr)
\bigl(1-v_{j}(t+s)\bigr) \bigr)
\\
&=& \E\bigl[v_{i}(t)v_{i}(t+s)\bigr]\prod
_{j<i} \biggl(1-\frac{2}{1+\theta}+\E \bigl[v_{j}(t)v_{j}(t+s)
\bigr] \biggr).
\end{eqnarray*}
Since $\operatorname{Corr}[v_{i}(t),v_{i}(t+s)]=\e^{-\lambda s}$ with
$\lambda
=(1+\theta)/2$ (cf.~Bibby, Skovgaard and S\o rensen \cite{BSS05}),
it follows that
$E[v_{i}(t)v_{i}(t+s)]=(1+\theta)^{-2}[1+(2+\theta)^{-1}\theta
\e^{-(1+\theta)s/2}]$, so
\begin{eqnarray*}
k_{s}&=& \bigl(c_{1}+c_{2}\e^{-(1+\theta)s/2}
\bigr) \sum_{i\ge1} \bigl(c_{1}
\theta^2 +c_{2}\e^{-(1+\theta)s/2} \bigr)^{i-1},
\\
c_{1}&=& \frac{1}{(1+\theta)^{2}},\qquad c_{2}=\frac{\theta
}{(1+\theta)^{2}(2+\theta)}
\end{eqnarray*}
from which the statement follows by direct computation.
Note that by
exchanging the limit operation with the sum and the integral (of
positive terms) in $\lim_{s\rightarrow0}k_{s}$, we obtain $\lim_{s\rightarrow0}k_{s}=(1+\theta)^{-1}$, hence $\lim_{s\rightarrow
0}\mathrm{Corr}(P_{t}(A),P_{t+s}(A))=1$.


\section{Algorithm details}\label{appa}

We illustrate the complementary details of the summary of the
simulation-based procedure, provided in Section~\ref{postcomputation},
for estimating the time-varying density which generates the data.
We will refer to quantities there introduced whenever this does not
compromise the readability.
Here we provide the general algorithm based on a GEM diffusive mixture
as discussed in Section~\ref{secdiscussion}. For ease of the reader,
the simplifications implied by choosing a Dirichlet diffusive mixture
are made explicit in the last part of the present section.

To recall briefly the relevant notation, let ${y}^{(n)}=(y_{t_1},\ldots
,y_{t_n})$ be the data points observed at times $(t_1,\ldots,t_n)$,
$v(\cdot)=((v_{1}(t),v_{2}(t),\ldots), t\ge0)$ with $v_j(\cdot)\sim
^{\mathrm{ind}} \operatorname{WF}(a_j,b_j)$ as in (\ref{wf-sde}),
$x=(x_1,x_2,\ldots)$
with $x_{j}\sim^{\mathrm{i.i.d.}}G$, for $G\in\P(\X)$ nonatomic. So, for example,
$\pi
_{v}$ in \eqref{lawv} becomes a $\operatorname{Beta}(a_{j},b_{j})$.
Note that the $(a_j,b_j)$ parameters must be chosen such that $\sum_{j\ge1} w_j(0)=1$, with $w_j(t)$ as in \eqref{stick-breaking-WF} (the
Wright--Fisher dynamics would then imply the same holds for all $t\ge
0$). See, for example, Ishwaran and James \cite{IJ01} for sufficient
conditions. Given the
discussion in Section~\ref{postcomputation}, it remains to make
explicit how to update the random measures locations and weights, the
slice and membership variables, and how to use the slice sampling on
the Wright--Fisher transition density. We treat these issues separately.

\textit{Updating the locations}.
Since the locations are not time dependent, these are updated as in
Kalli, Griffin and Walker \cite{KGW11}. That is
%
\begin{equation}
\label{nodepxs} \mathcal{L}(\msf{x}_j\mid\cdots)\propto G(
\msf{x}_j) \prod_{\{i: s_i
= j\}}
K(y_{t_i}\mid\msf{x}_j)
\end{equation}
for $j=1,\ldots,\msf{m}$, so that only a finite number of locations
need to be sampled.

\textit{Updating the weights}.
We need the full conditional distributions for each of the $m\times n$
Wright--Fisher values $v_{j}(t_{i})$, where $j=1,\ldots,m$, $m$ is as
in (\ref{m=max}) and $i=1,\ldots,n$.
Hence, for each $j=1,\ldots,m$ we have
%
\begin{eqnarray}
\label{updateV} %
\mathcal{L}\bigl(\msf{v}_j(t_1)
\mid\cdots\bigr)&\propto& \msf{p}_{\msf
{v}}\bigl(\msf {v}_j(t_{2})
\mid\msf{v}_j(t_1) \bigr) \msf{\pi}_{\msf{v}}\bigl(
\msf{v}_j(t_{1}) \bigr) \msf{v}_j(t_1)^{\bb{I}(s_1=j)}
\bigl(1-\msf{v}_j(t_1)\bigr)^{\bb
{I}(s_1>j)},
\nonumber
\\
\mathcal{L}\bigl(\msf{v}_j(t_i)\mid\cdots\bigr)&\propto&
\msf{p}_{\msf
{v}}\bigl(\msf {v}_j(t_{i+1})\mid
\msf{v}_j(t_i) \bigr) \msf{p}_{\msf{v}}\bigl(\msf
{v}_j(t_{i})\mid\msf{v}_j(t_{i-1})
\bigr) \msf{v}_j(t_i)^{\bb{I}(s_i=j)}
\nonumber
\\[-8pt]
\\[-8pt]
\nonumber
&&{}\times \bigl(1-
\msf{v}_j(t_i)\bigr)^{\bb{I}(s_i>j)},
\qquad i\neq1,n,
\\
\mathcal{L}\bigl(\msf{v}_j(t_n)\mid\cdots\bigr)&\propto&
\msf{p}_{\msf
{v}}\bigl(\msf {v}_j(t_{n})\mid
\msf{v}_j(t_{n-1}) \bigr) \msf{v}_j(t_n)^{\bb{I}(s_n=j)}
\bigl(1-\msf{v}_j(t_n)\bigr)^{\bb{I}(s_n>j)}.\nonumber %
\end{eqnarray}
Note that dropping the dependence on time in the weights processes
$w_{j}(t)$ would yield
\begin{eqnarray*}
%
\mathcal{L}(\msf{v}_j\mid
\cdots)&\propto& \pi_{\msf{v}}(\msf{v}_j) \prod
_{\{i: s_i = j\}} \biggl[\msf{v}_{j}\prod
_{k<j}(1-\msf {v}_k) \biggr]
\\
&\propto& \pi_{\msf{v}} (\msf{v}_j) \msf{v}_{j}^{\sum_{i=1}^n\bb
{I}(s_i=j)}
(1-\msf{v}_{j})^{\sum_{i=1}^n\bb{I}(s_i>j)}. 
\end{eqnarray*}
Letting $\pi_{\msf{v}_j}=\operatorname{Beta}(a_j,b_j)$, the previous
simplifies to
\[
\mathcal{L}(\msf{v}_j\mid\cdots)=\operatorname{Beta}
\Biggl(a_j+\sum_{i=1}^n\bb
{I}(s_i=j),b_j+\sum_{i=1}^n
\bb{I}(s_i>j)\Biggr),
\]
which is the usual posterior update in the framework of stick-breaking
random probability measures based on Beta distributed stick-breaking
component. See Kalli, Griffin and Walker \cite{KGW11}.

\textit{Updating the slice and membership variables.}
For each $i=1,\ldots,n$ we have
\[
\mathcal{L}(u_{t_i}\mid\cdots)=\msf{U}(0,\psi_{s_{t_i}})
\]
and
\[
\mathcal{L}(s_{t_i}\mid\cdots)\propto\frac{\msf
{w}_{s_{t_i}}(t_i)}{\psi_{s_{t_i}}} K
\bigl(y_{t_i}\mid\msf {x}_{s_{t_i}}(t_i)\bigr) \bb{I}
\bigl(s_{t_i}\in \{k\dvt \psi _{s_{t_i}}>u_{t_i} \}\bigr).
\]
Note that since $  \{k\dvt \psi_{s_{t_i}}>u_{t_i} \}$ is a
finite set,
the above distribution involves a finite sampling, namely from
$s_{t_i}=1,\ldots,\lfloor\psi^*(u_{t_i})\rfloor$, where $\psi^*$
denotes the inverse of $\psi$.

\textit{Slicing the Wright--Fisher transition density}.
The analytic form for the transition density of the Wright--Fisher
diffusion model does not take a simple closed form. One could attempt
to use the corresponding spectral representation, which involves an
infinite series of orthogonal Jacobi polynomials. 
However, such infinite number of arguments with alternating sign
prevents a robust evaluation of the transition density, especially for
the extreme points of the state space. Instead, here we opt to use a
slightly more general representation given in Mena and Walker \cite
{MW09}. That is, an
equivalent formulation of the transition density of the Wright--Fisher
diffusion is
%
\begin{equation}
\label{transWF} \msf{p}_{\msf{v}}\bigl(\msf{v}(t)\mid\msf{v}(0)\bigr)=\sum
_{m=0}^\infty r_{t}(m) D\bigl(
\msf{v}(t)\mid m,\msf{v}(0)\bigr),
\end{equation}
where
%
\begin{equation}
\label{funcr} r_t(m)=\frac{(a+b)_m \e^{-mct}}{m!}\bigl(1-\e^{-ct}
\bigr)^{a+b}
\end{equation}
and
%
\begin{equation}
\label{transWFb} D\bigl(\msf{v}(t)\mid m,\msf{v}(0)\bigr)=\sum
_{k=0}^m \operatorname {Beta}\bigl(\msf{v}(t)|
a+k,b+m-k\bigr)\operatorname{Bin}\bigl(k| m, \msf{v}(0)\bigr).
\end{equation}
Here $\operatorname{Beta}(\cdot\mid a,b)$ is the Beta density with parameters
$a,b$ and $\operatorname{Bin}(\cdot\mid m,q)$ is the Binomial probability
function with $m$ trials and success probability $q$.
A reparametrization of (\ref{wf-sde}) given by letting $c=(a+b-1)/2$
leads to writing
\[
\mathrm{ d}v(t)= \biggl( \frac{c(a-(a+b)v(t))}{a+b-1} \biggr) \,\mathrm{ d}t+ \biggl(
\frac
{2c}{a+b-1}v(t) \bigl(1- v(t)\bigr) \biggr)^{1/2}\,\mathrm{ d}B(t).
\]
%
The above representation is valid for $a+b>1$, in which case $0$ and
$1$ are entrance boundaries. Such condition rules out the inconvenient
case of weights $w_{j}(t)$ in \eqref{stick-breaking-WF} become $0$ or $1$.
A byproduct of the above re-parametrization is that it makes explicit
the rate of decay in the autocorrelation function of the Wright--Fisher
diffusion, which for the Dirichlet process case \eqref
{stick-breaking-WF} reduces to $\operatorname
{Corr}[v_{j}(t),v_{j}(t+s)]=\e^{-(1+\theta)s/2}$. See, for example,
Bibby, Skovgaard and S\o rensen~\cite{BSS05}.

The representation \eqref{transWF} is appealing not only due to the
fact that it involves elementary functions, but also and foremost
since, unlike in the spectral decomposition, the summands are all
positive. It follows that truncations, or rather random truncations
such as those invoked by the slice method, are feasible.
Indeed, with techniques similar to those used for \eqref{dnpmaug2} we
can augment the transition density in order to avoid the infinite
computation. Introduce then $(o_j,k_j,d_j)$ such that
\begin{eqnarray*}
&&\msf{p}^{\msf{v}}_t\bigl(\msf{v}_j(t),
o_j, k_j, d_j\mid\msf{v}_j(0)
\bigr)
\\
&&\quad= \bb{I}\bigl(o_j<g(d_j)\bigr)\frac{r_{j,t}(d_j)}{g(d_j)}
\operatorname {Beta}\bigl(\msf {v}_j(t)| a_j+k_j,b_j+d_j-k_j
\bigr) \operatorname{Bin}\bigl(k_j| d_j,
\msf{v}_j(0)\bigr),
\end{eqnarray*}
where as before $d\mapsto g(d)$ is a decreasing function with known
inverse $g^*$ and $r_{j,t}(\cdot)$ denotes \eqref{funcr} computed on
parameters $(a_j,b_j,c_j)$.
Augmenting for $n$ observations by means on $(o^i_j, k^i_j,
d^i_j)_{i=2}^n$ leads to the likelihood for $\msf{m}$ processes
\begin{eqnarray*}
&&\mathcal{L}\bigl(\msf{v}_\msf{1:m} ^{(n)},
\msf{o}_\msf{1:m}^{(n)},\msf {k}_\msf{1:m}^{(n)},
\msf{d}_\msf{1:m}^{(n)}\mid a_j,b_j,c_j
\bigr)
\\
&&\quad=\prod_{j=1}^\msf{m}\pi_{\msf{v}}
\bigl(\msf{v}_j(t_1)\bigr)\prod
_{i=2}^n \bb{I}\bigl(o^i_j<g
\bigl(d^i_j\bigr)\bigr)\frac{r_{j,\tau_i}(d^i_j)}{g(d^i_j)}
\\
&&\hspace*{72pt}\qquad{} \times\operatorname{Beta}\bigl(\msf{v}_j(t_i)|
a_j+k^i_j,b_j+d^i_j-k^i_j
\bigr)\operatorname {Bin}\bigl(k^i_j|
d^i_j, \msf{v}_j(t_{i-1})\bigr),
\end{eqnarray*}
where $\tau_i=t_{i}-t_{i-1}$, and the subscript ``$1:m$'' is
interpreted as in \eqref{1mnotation}. Therefore, given a prior $\pi
(a_j,b_j,c_j)$, the posterior distribution for $(a_j,b_j,c_j)$ is
%
\begin{equation}
\label{postparam} \mathcal{L}(a_j,b_j,c_j
\mid\cdots)\propto\mathcal{L}\bigl(\msf {v}_\msf {1:m} ^{(n)},
\msf{o}_\msf{1:m}^{(n)},\msf{k}_\msf{1:m}^{(n)},
\msf {d}_\msf{1:m}^{(n)}\mid a_j,b_j,c_j
\bigr) \pi(a_j,b_j,c_j).
\end{equation}
Furthermore, the full conditionals for the latent variables $(o^i_j,
k^i_j, d^i_j)_{i=2}^n$ for each $j=1,\ldots,\msf{m}$ are given by
$\mathcal{L}({o_j^i}\mid\cdots)=\msf{U}(o_j^i\mid0,g({d_j^i}))$,
\begin{eqnarray*}
\mathcal{L}\bigl({k_j^i}\mid\cdots\bigr)
\pmatrix{d_j^i\cr k_j^i}
\frac
{\bb
{I}(k_j^i\in\{0,\ldots,d_j^i\})}{\Gamma(a_j+k_j^i)\Gamma
(b_j+d_j^i-k_j^i)} \biggl\{\frac{\msf{v}_j(t_i)\msf{v}_j(t_{i-1})}{
(1-\msf
{v}_j(t_i))(1-\msf{v}_j(t_{i-1}))} \biggr\}^{k^i_{j}}
\end{eqnarray*}
%
and
\[
\mathcal{L}\bigl({d_j^i}\mid\cdots\bigr)\propto
\frac{\Gamma
(a_j+b_j+d_j^i)^2
[(1-\msf{v}_j(t_{i}))(1-\msf{v}_j(t_{i-1}))
]^{d_j^i}}{\e^{d_j^ic_j\tau
_i}\Gamma(b_j+d_j^i-k_j^i)\Gamma(d_j^i-k_j^i+1)g(d_j^i)} \bb {I}\bigl(k_j^i\leq
d_j^i\leq g^*\bigl(o_j^i\bigr)
\bigr).
\]

The supports of $\mathcal{L}({k_j^i}\mid\cdots)$ and $\mathcal
{L}({d_j^i}\mid\cdots)$ are discrete and bounded, so sampling from such
distributions is straightforward, for example, via the inverse
cumulative distribution function method.

\textit{Gibbs sampler for diffusive DP mixtures}. 
When instead of a diffusive GEM mixture one chooses the special case of
a diffusive Dirichlet process mixture, this corresponds to letting
$a_j=1$, $b_j=\theta$ and $c_j=c$ for all $j=1,2,\ldots$ in the above
derivation, or equivalently to take independent and identically
distributed Wright--Fisher processes as in (\ref{stick-breaking-WF}).
With these specifications the full conditionals \eqref{updateV} for the
weights processes simplify to
\begin{eqnarray*}
\cl{L}\bigl(\msf{v}_j(t_1)
\mid\cdots\bigr)&=&\operatorname {Beta}\bigl(\msf{v}_j(t_1)
\mid1+k_j^2+\bb{I}(s_1=j); \theta
+d^2_j-k^2_j+\bb
{I}(s_1>j)\bigr),
\\
\cl{L}\bigl(\msf{v}_j(t_i)\mid\cdots\bigr)&=&
\operatorname {Beta} \bigl(\msf{v}_j(t_i)
\mid1+k_j^i+k_j^{i+1}+
\bb{I}(s_i=j);\\
&&\hspace*{24pt}\theta+d_j^i+d_j^{i+1}-k_j^i-k_j^{i+1}+
\bb{I}(s_i>j) \bigr),
\qquad i=2,\ldots,n-1,
\\
\cl{L}\bigl(\msf{v}_j(t_n)\mid\cdots\bigr)&=&
\operatorname {Beta}\bigl(\msf{v}_j(t_n)
\mid1+k_j^n+\bb{I}(s_n=j); \theta
+d^n_j-k^n_j+\bb
{I}(s_n>j)\bigr).
\end{eqnarray*}
Assuming independent priors for $\theta$ and $c$, and taking the
logarithm, the full conditionals \eqref{postparam} become
\begin{eqnarray*}
\log\cl{L}(\theta\mid\cdots) &\propto& \log\bigl(\pi(\theta)\bigr)-\msf{m}(n-2)
\log\Gamma(1+\theta )-\msf{m}\log \Gamma(\theta)+\msf{m}\theta\sum
_{i=2}^n\log\bigl(1-\e^{-c\tau_i}\bigr)
\\
&&{}+\theta\sum_{j=1}^{\msf{m}}\sum
_{i=1}^{n}\log\bigl(1-\msf {v}_j(t_i)
\bigr)+2\sum_{j=1}^{\msf{m}}\sum
_{i=2}^{n} \log\Gamma\bigl(1+\theta+d_j^i
\bigr)
\\
&&{}-\sum_{j=1}^{\msf{m}}\sum
_{i=2}^{n} \log\Gamma\bigl(\theta+d^i_j+k_j^i
\bigr)
\end{eqnarray*}
and
\[
\log\cl{L}(c\mid\cdots)\propto\log\bigl(\pi( c )\bigr)+\msf{m}(1+\theta )\sum
_{i=2}^n \log\bigl(1-\e^{-c\tau_i}
\bigr)-c\sum_{j=1}^{\msf{m}}\sum
_{i=2}^{n}d_j^i
\tau_i.
\]
These can be sampled using the Adaptive Rejection Metropolis Sampling
(ARMS) algorithm.
\end{appendix}

\section*{Ackowledgements}

{The authors are grateful to an Associate Editor and two anonymous referees for
carefully reading the manuscript and for providing constructive
comments which led to an improvement of the presentation.}
The first author was partially supported by \emph{Consejo Nacional de
Ciencia y Tecnolog\'ia de M\'exico} project 131179, during the
elaboration of this work.
The second author is supported by the European Research Council (ERC)
through StG ``N-BNP'' 306406.

%





\printhistory
\end{document}